\documentclass[conference]{IEEEtran}
\usepackage[tbtags]{amsmath}
\usepackage{amssymb}
\usepackage{subfigure}
\usepackage{epsfig}

\newtheorem{thm}{Theorem}[section]

\newcommand{\bit}{\begin{itemize}}
\newcommand{\eit}{\end{itemize}}
\newcommand{\beq}{\begin{equation}}
\newcommand{\eeq}{\end{equation}}
\newcommand{\beqn}{\begin{equation*}}
\newcommand{\eeqn}{\end{equation*}}
\newcommand{\bea}{\begin{eqnarray}}
\newcommand{\eea}{\end{eqnarray}}
\newcommand{\bean}{\begin{eqnarray*}}
\newcommand{\eean}{\end{eqnarray*}}

\begin{document}

\title{On approximating Gaussian relay networks with deterministic networks}

\author{M. Anand and P. R. Kumar \\
\authorblockA{Dept. of ECE and CSL,
Univ. of Illinois, Urbana, IL 61801, USA\\
Email: \{amurali2, prkumar\}@illinois.edu}
\thanks{This material is
based upon work partially supported by USARO under Contract Nos.
W911NF-08-1-0238 and W-911- NF-0710287, and NSF under Contract Nos.
ECCS-0701604, CNS-07-21992, CNS-0626584, and CNS-05-19535.}}

\maketitle

\begin{abstract}
We examine the extent to which Gaussian relay networks can be
approximated by deterministic networks, and present two results, one
negative and one positive.

The gap between the capacities of a Gaussian relay network and a
corresponding linear deterministic network can be unbounded. The key
reasons are that the linear deterministic model fails to capture the
phase of received signals, and there is a loss in signal strength in
the reduction to a linear deterministic network.

On the positive side, Gaussian relay networks are indeed well
approximated by certain discrete superposition networks, where
the inputs and outputs to the channels are discrete, and channel
gains are signed integers.

As a corollary, MIMO channels cannot be approximated by the linear
deterministic model but can be by the discrete superposition model.
\end{abstract}

\section{Introduction}

There have been many efforts to determine the capacities of Gaussian
networks with multiple sources and destinations.  A recent proposal
is to approximate a given Gaussian network by a linear deterministic
model which is noise-free, linear, and easy to analyze. This model
was introduced in \cite{ADT1, ADT2} where the capacity of linear
deterministic networks with a single source-destination pair is
determined. This approach was successful for certain Gaussian
networks like the interference channel \cite{BT}, and the MAC and
broadcast networks \cite{ADT2}, where the gap between the capacities
of the Gaussian network and the linear deterministic network is
bounded by a constant independent of channel gains. Most subsequent
research on the linear deterministic model has been focussed on
deriving coding schemes for Gaussian networks that are inspired by
those for the deterministic network \cite{AST,CJS}.

\subsection{Our results}

We consider Gaussian relay networks with a single source-destination
pair and multiple relay nodes.  The relays have no data to transmit
but help the source in sending its data to the destination. We
analyze the extent to which a linear deterministic model can approximate
such networks, by comparing their capacities.  We show that the gap
in the capacities can be unbounded.  This is since the linear deterministic
model cannot capture the phase of a channel gain.  Even restricted
to Gaussian networks with positive channel gains, the linear
deterministic model is not a good approximation.

As a positive result towards approximating Gaussian networks, we
show that an earlier {\em discrete superposition} model with
discrete inputs and outputs \cite{BT} serves as a good approximation
for Gaussian relay networks.

A corollary is MIMO channels cannot be approximated by linear
deterministic model, but can be by discrete superposition
models.

\section{Preliminaries} \label{sec:prelim}

\subsection{Model} \label{sec:model}

We consider a wireless network represented as a directed graph
$(\mathcal{V}, \mathcal{E})$, where $\mathcal{V} = \{0, 1, \ldots,
M\}$ represents nodes, and the directed edges in $\mathcal{E}$
correspond to wireless links.  Denote by $h_{ij}$ the complex number
representing the fixed channel gain for link $(i,j)$.  Let the
complex number $x_i$ denote the transmission of node $i$.  Every
node has an average power constraint, taken to be $1$.  Node $j$
receives \beqn
    y_j = \sum_{i \in \mathcal{N}(j)} h_{ij} x_{i} + z_{j} ,
\eeqn where $\mathcal{N}(j)$ is the set of its neighbors and $z_j$ is
complex white Gaussian noise, $\mathcal{CN}(0,1)$, independent of
the transmitted signals.

\subsection{Constructing the linear deterministic network}
\label{sec:det_const}

Perhaps the best way to understand the linear deterministic model
\cite{ADT1} is to develop it in a point-to-point setting.
Consider a simple AWGN channel \beqn y \ = \ hx + z \eeqn with capacity
$C \ = \ \log(1 + |h|^2)$.  Let $C_D \ := \ \lfloor \log |h|^2
\rfloor $ approximately denote its capacity in the high SNR regime.  We construct a deterministic network of capacity $C_D$ with a source that transmits a binary vector $x$ (of length at least $C_D$ with bits ordered from left to right) and a channel that attenuates the signal by allowing $C_D$ most significant bits to be received at the destination.

In a general Gaussian network, choose all the inputs and outputs of channels to be binary vectors of length
$\max_{(i,j) \in \mathcal{E}} \lfloor \log |h_{ij}|^2 \rfloor$.  Each link with channel gain $h$ is
replaced by a matrix that shifts the input vector and allows $\lfloor
\log |h|^2 \rfloor$ most significant bits of the input to pass through.  At a
receiver, shifted vectors from multiple inputs are added bit by bit over
the binary field. This models the partially destructive nature of
interference in wireless.  The channel is simply a linear
transformation over the binary field. Modeling the broadcast feature
of wireless networks, a node transmits the same vector on all
outgoing links, albeit with different attenuation and
the number of significant bits arriving at a receiver depends on the
channel gain.

\begin{figure}[ht]
    \subfigure[A Gaussian network.]{
    \label{fig:awgn_mimo}
    {\includegraphics[width=1in, height = 0.7in]{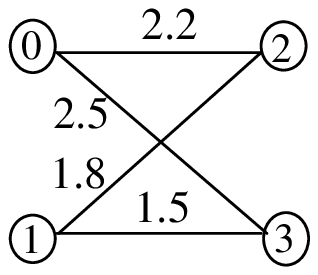}}
    } \hfill
    \subfigure[Linear deterministic model.]{
    \label{fig:det_mimo}
    {\includegraphics[width=1.4in, height = 0.7in]{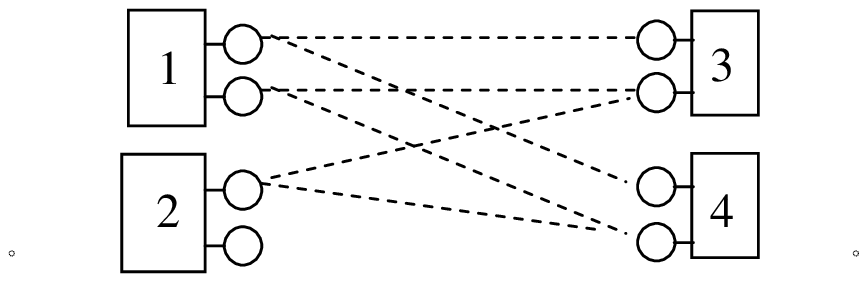}}}
    \caption{Linear deterministic model of Gaussian network.
    (Note $\lfloor \log 2.2^2 \rfloor = \lfloor \log 2.5^2 \rfloor = 2$,
    $\lfloor \log 1.8^2 \rfloor = \lfloor \log 1.5^2 \rfloor = 1$)}
    \label{fig:det_nw_mimo}
\end{figure}

The capacity of linear deterministic
relay networks  with a single source-destination pair is known \cite{ADT2}. Recently the capacity of
Gaussian networks with a single source-destination pair was
approximately computed \cite{ADT3}. Therefore we can compare them.

\subsection{Cut-set bound}

The cut-set bound \cite{Gam} of a single source-destination pair
network, with source $0$ and destination $M$, is \beq
    C \ \leq \ \max_{p(x_0, x_1, \ldots, x_{M-1})} \ \min_{\Omega \in
    \Lambda} \
    I(X_{\Omega}; Y_{\Omega^c} | X_{\Omega^c} )  ,
    \label{eq:cs_bd}
\eeq where $C$ is capacity of the network $\Lambda$ is the set of all
partitions of $\mathcal{V}$ with $0 \in \Omega$ and $M \in
\Omega^c$.

\subsubsection{Cut-set bounds for linear deterministic networks}
Since outputs are a function of inputs in a linear
deterministic network, $I(X_{\Omega} ; Y_{\Omega^c} | X_{\Omega^c})
\ = \ H(Y_{\Omega^c} |  X_{\Omega^c})$.  Maximum value of the mutual information equals the
rank of the transfer matrix $\mathcal{G}_{\Omega, \Omega^c}$
associated with the cut $\Omega$ \cite{ADT2}, where the rank is
determined over an appropriate finite field.  An optimal input distribution is
input variables independent and uniformly distributed over the underlying field.  Hence \eqref{eq:cs_bd} simplifies to \beq
    C \ \leq \ \min_{\Omega} \ \text{rank} \ \mathcal{G}_{\Omega, \Omega^c}.
    \label{eq:cap_det}
\eeq  For Fig.~\ref{fig:det_mimo}, rank of $\mathcal{G}_{\Omega,
\Omega^c}$ over $\mathbb{F}_2$, with $\Omega = \{0, 1\}$, is $2$ with
\beq
    \mathcal{G}_{\Omega, \Omega^{c}} \ = \left[ \begin{array}{cc}
        \left[ \begin{array}{cc}
        1 & 0 \\ 0 & 1 \end{array} \right] &
        \left[\begin{array}{cc}
        0 & 0 \\ 1 & 0 \end{array} \right] \\
        \left[\begin{array}{cc}
        1 & 0 \\ 0 & 1 \end{array} \right] &
        \left[\begin{array}{cc}
        0 & 0 \\ 1 & 0 \end{array} \right]
        \end{array} \right]. \nonumber
\eeq
The cut-set bound in \eqref{eq:cap_det} is achieved by random linear coding and hence is the capacity of the network \cite{ADT2}.

\subsubsection{Cut-set bounds for Gaussian Networks}
\label{sec:gauss_cs_bd}
Choosing $\{X_j\}$'s to be i.i.d. $\mathcal{CN}(0,1)$ weakens the
bound in \eqref{eq:cs_bd} by at most $(M + 1)$ bits for any choice
of channel gains and yields \beq
    \overline{C} \ := \ \min_{\Omega \in
\Lambda} \
    \log | I + \mathcal{H}_{\Omega} \mathcal{H}_{\Omega}^{\dag}| \label{eq:cs_bd_logdet} ,
\eeq where $\mathcal{H}_{\Omega}$ is the transfer matrix of the MIMO
channel corresponding to cut $\Omega$ and $|.|$ is determinant of the matrix
\cite{ADT3}.

In \cite{ADT3}, a coding scheme is developed that achieves a rate no
less than $\overline{C} - \kappa$ for all channel gains, with
$\kappa$ only depending on the number $M$ of nodes.  Nodes quantize
and forward their data and the destination eventually decodes the
transmitted symbol after hearing from all the nodes. This scheme is
inspired by a coding scheme developed in
\cite{ADT2} for a class of general deterministic networks.

\section{Unbounded capacity difference of linear deterministic model} \label{sec:neg}

We will show that capacity of the linear deterministic network
can be much lower than that of the original Gaussian network with
their difference unbounded as channel gains are varied.

From the previous section, if capacities of the linear
deterministic network differed and the Gaussian network differed by
a bounded amount, then the difference between their individual
cut-set bounds would also be so.  We establish unboundedness of the
difference by comparing mutual information across cuts in a
Gaussian network with ranks of the corresponding cuts in the
linear deterministic network.

For the rest we choose the inputs $X_i$ for the Gaussian network to
be i.i.d. $\mathcal{CN}(0,1)$, noting that this can achieve the maximum mutual information across any cut within a constant bound. For
the linear deterministic network, we choose inputs that are
independent and uniformly distributed over their range since that
maximizes mutual information across a cut.

\subsection{Counterexample to constant bit approximation} \label{sec:counter}

\begin{figure}[ht]
    {\subfigure[Gaussian network.]{
    \label{fig:ortho_gauss}
    \includegraphics[width=1.5in]{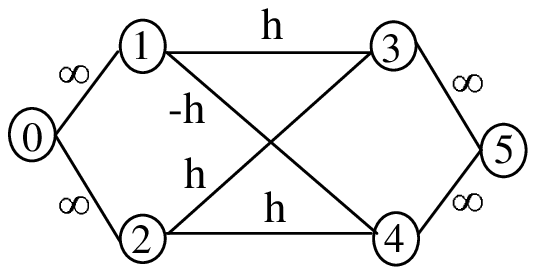}
    }}
    {\subfigure[Portion of deterministic network.]{
    \label{fig:por_gauss}
    \includegraphics[width=1.5in, height = 0.7in]{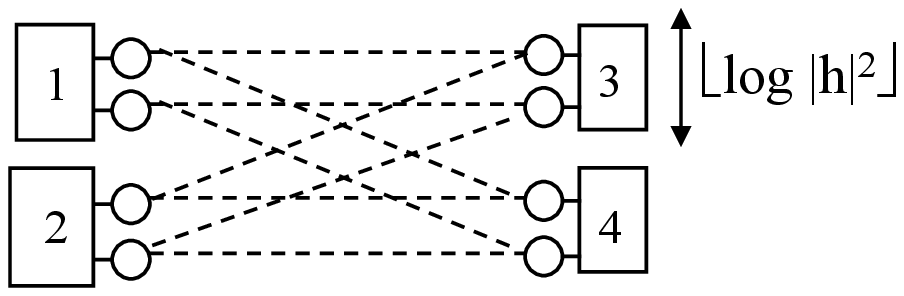}}}
    \caption{Counterexample.}
    \label{fig:counter_gauss}
\end{figure}

Consider the network in Fig.~\ref{fig:ortho_gauss} where the channels
marked as $\infty$ have very high capacity.  The mutual information
across $\Omega = \{ 0, 1, 2\}$ is \bean
    I(X_{\Omega}; Y_{\Omega^{c}} | X_{\Omega^{c}}) & = & \log | I + \mathcal{H} \mathcal{H}^{\dag} | \nonumber\\
    & = & 2 \log (1 + 2 |h|^2) \nonumber\\
    & = & 4 \log |h| + O(1), \ \text{as } |h| \rightarrow \infty,
\eean with $
    \mathcal{H} \ = \ \left[ \begin{array}{cc}
                h & -h \\
                h & h
                \end{array}\right]. \nonumber
$ This is the minimum among all cuts and is therefore the capacity.

In the corresponding linear deterministic network, the transfer
matrix of $\Omega \ = \ \{0, 1, 2\}$ in Fig.~\ref{fig:por_gauss} is
$\mathcal{G} \ = \ \left[ \begin{array}{cc}
                I & I \\
                I & I
                \end{array}\right]$, where each identity matrix has dimension $\lfloor \log |h|^2
\rfloor$. The capacity of the network is the rank of $\mathcal{G}$, i.e., $\lfloor 2 \log |h|
\rfloor$.

The gap between capacities of the Gaussian network in
Fig.~\ref{fig:ortho_gauss} and its deterministic counterpart is $2
\log |h| + O(1)$.  Therefore the gap cannot be bounded independently of
channel gains.

The linear deterministic model considers only the magnitude of a channel
gain and fails to capture its phase.  Constructing the deterministic
model over a larger prime field does not help either.

\subsection{Gaussian networks with positive channel gains}
\label{sec:zero_phase}

Unfortunately phase is not the only problem.  We construct a
Gaussian network with {\em positive} channel coefficients that
cannot be approximated by a linear deterministic network.

\begin{figure}[ht]

    {\subfigure{
    \label{fig:pos_ch}
    \includegraphics[width=1.5in]{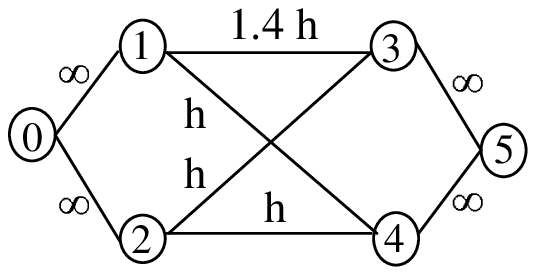}}} \hfill
    {\subfigure{
    \label{fig:pos_ch_q}
    \includegraphics[width=1.5in]{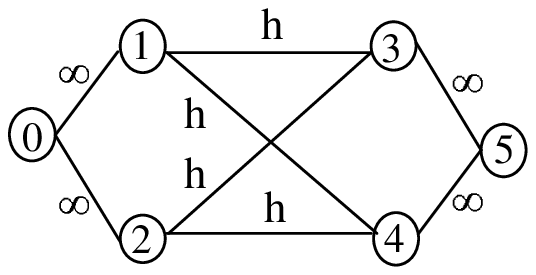}}}
    \caption{Two networks with positive channel coefficients. Both have the same linear deterministic network in Fig.~\ref{fig:por_gauss} as their counterpart.}
    \label{fig:pos_net}
\end{figure}

Consider the Gaussian networks in Fig.~\ref{fig:pos_net}, where $h =
2^k$ for $k \in \mathbb{Z}_{+}$. The linear deterministic network
corresponding to both Gaussian networks is the same.  However, the
difference in the capacities of the Gaussian networks is unbounded.

The capacity of the network in Fig.~\ref{fig:pos_ch} is \bea
    C & = & I(X_{\Omega}; Y_{\Omega}^{c} | X_{\Omega}^{c}) + O(1) \nonumber\\
    & = & 4 \log h + O(1) \ \ (\text{as} \ h \rightarrow
    \infty), \nonumber
\eea with $\Omega = \{0, 1, 2\}$, while the capacity of the network in
Fig.~\ref{fig:pos_ch_q} is \bea
    \hat{C} & = & I(X_{\Omega}; Y_{\Omega}^{c} | X_{\Omega}^{c}) + O(1) \nonumber\\
    & = & 2 \log h + O(1) \ \ (\text{as} \ h \rightarrow
    \infty). \nonumber
 \eea
So difference in capacities of the linear deterministic
network in Fig.~\ref{fig:por_gauss} and at least one of its counterparts in Fig.
\ref{fig:pos_net} must be unbounded in $h$.

One may wonder if taking the channel gains into account and
quantizing the gains with respect to a field larger than
$\mathbb{F}_2$ will provide a bounded error approximation. However
this reasoning is flawed since then the gap in the capacities would
be a function of the chosen prime and thus, in turn, a function of
channel gains.

\section{Approximating Gaussian networks by discrete superposition networks}
\label{sec:dis_det}

A question arises if there is any other deterministic model
for approximating Gaussian networks.  We now show that an alternate
deterministic model, first mentioned in \cite{BT}, is indeed a good
approximation of a Gaussian relay network.  This model, which we
call a {\em discrete superposition} model, captures the phases
of channel gains, and ensures that the signal strength does not drop
due to quantization of channel gains.

\subsection{The discrete superposition model}

We define the inputs, outputs, and channels in a discrete
superposition network.  Let \beqn n \ := \ \max_{(i,j) \in
\mathcal{E}} \max \{ \lfloor \log |h_{ijR}| \rfloor, \lfloor \log
|h_{ijI}| \rfloor\} , \eeqn where $h_{ij} = h_{ijR} + \imath
h_{ijI}$.  The inputs are complex valued and both real and imaginary
parts can take $2^n$ equally spaced discrete values from $\{0,
2^{-n}, \ldots, 1 - 2^{-n}\}$.  It helps to think of either the real
or imaginary part of an input in terms of its binary representation,
i.e., $(x = 0.x(1) x(2) \dots x(n))$ with each $x(i) \in
\mathbb{F}_2$.

The real and imaginary parts of channel gains are quantized to
integers by neglecting their fractional parts.  The channel between
two nodes multiplies the complex input by the corresponding channel
gain, and then truncates it by neglecting the fractional components
of both real and imaginary parts of the product.  The outputs of all
incoming channels at a receiver node are complex numbers with
integer real and imaginary parts.  All the outputs are added up at a
receiver by standard summation over $\mathbb{Z} + \imath
\mathbb{Z}$.

This model retains the essential superposition property of the
channel. The truncation of channel coefficients does not
substantially change the channel matrix in the high SNR limit. Also,
the effect of noise is captured in essentially the same way as in
the linear deterministic model by truncating least significant bits.

This discrete superposition model was first used in \cite{BT} in a
sequence of networks that reduced the Gaussian interference channel
to a linear deterministic network.  We use some of the techniques
from \cite{BT} in the proof below.  In \cite{ADT2}, it was shown
that the cut-set bound is achievable for such deterministic
networks, provided attention is restricted to product distributions
for the input signals.  The main result presented below entails
showing that the loss in restricting attention to product
distributions for inputs is bounded for discrete superposition relay
networks.  The connection with Gaussian relay networks or with MIMO
channels (see the following Sec. \ref{sec:mimo}) was not made in
\cite{BT}.  There is also no cooperation among the input or output
nodes in \cite{BT}, which is a key ingredient of the proof below.

\vspace*{0.05in}

\begin{thm}
\label{thm:det_gap} The difference in capacities of a Gaussian
relay network with a single source-destination pair and the
corresponding discrete superposition network is bounded with the
bound depending only on the number of nodes.
\end{thm}
\noindent \begin{proof}  \noindent We prove the result by assuming
that the network permits only real valued signals.  Extending the
result to Gaussian networks with complex valued signals is
straightforward, but involves more bookkeeping as noted at the end.

We show that achievable rates in the Gaussian network are within a
bounded gap of achievable rates in the discrete superposition
network and vice versa.

First we start with the Gaussian network and reduce it to a
discrete superposition network in stages, bounding the loss in
mutual information at each stage.  For specificity consider a cut
with two input nodes and two output nodes. Let the output signals be
\beq
    y_i = h_{i1} x_1 + h_{i2} x_2 + z_i, \ i = 1, 2.
    \label{eq:start}
\eeq Choose $x_1$, $x_2$ as i.i.d. $\mathcal{N}(0,1)$ (since this
choice is approximately optimal in the sense that it maximizes
mutual information of the cut up to a constant, see
Sec.~\ref{sec:gauss_cs_bd}). The mutual information across this cut
is \beq
    I(X_1, X_2; Y_1, Y_2) \ = \ \frac{1}{2} \, \log |I + \mathcal{H} \mathcal{H}^{t}| ,
    \label{eq:mi_two}
\eeq where $\mathcal{H}$ is the channel transfer matrix.  Our goal
is to derive inputs to the corresponding discrete superposition
channel from $x_1$ and $x_2$ such that mutual information of the
deterministic channel differs by no more than a constant from
\eqref{eq:mi_two}.

We begin by scaling all channel gains by half: \bea
    y_i^{'} & := & (h_{i1}/2) x_1 + (h_{i2}/2) x_2 + z_i \nonumber\\
    & = & h_{i1}^{'} x_1 + h_{i2}^{'} x_2 + z_i, \ i = 1, 2,
    \label{eq:scale}
\eea with $h_{ij}^{'} := h_{ij}/2$.  The mutual information
decreases by at most $1$ bit in comparison to \eqref{eq:mi_two},
since \beqn
    (I + \mathcal{H} \mathcal{H}^{t}) \ \preceq \
    4(I + \frac{\mathcal{H}}{2} \frac{\mathcal{H}^{t}}{2}) \eeqn as positive definite matrices. So,
    \beqn
    I(X_1, X_2; Y_1, Y_2) \ \leq \ I(X_1, X_2; Y_1^{'}, Y_2^{'}) +
    1.
\eeqn

Each $x_i$ can be split into its integer part $\hat{x}_i$ and fractional part $\hat{x}_i$.   More precisely $\hat{x}_i := (\text{sign
} x_i)) \lfloor |x_i| \rfloor$, $\bar{x}_i := x_i - \hat{x}_i$.  We discard $\hat{x}_i$ and retain $\bar{x}_i$.  Since $x_i$ satisfies unit average power constraint, $\bar{x}_i$ satisfies unit peak power constraint.
Define \beq
    \bar{y}_i \ := \ h_{i1}^{'} \bar{x}_1 + h_{i2}^{'} \bar{x}_2 + z_i, \ i = 1,
    2. \label{eq:quant}
\eeq Denote the discarded portion of the received signal by \beq
    \hat{y}_i = h_{i1}^{'} \hat{x}_1 + h_{i2}^{'} \hat{x}_2 , \ i = 1,
    2. \label{eq:discard}
\eeq Comparing with the channel in \eqref{eq:scale}, we get \bea
    \lefteqn{I(X_1, X_2; Y_1^{'}, Y_2^{'})} \nonumber\\
    &  \leq & I(X_1, X_2; Y_1^{'}, Y_2^{'}, \hat{Y}_1, \hat{Y}_2, \bar{Y}_1, \bar{Y}_2) \nonumber \\
    & = & I(X_1, X_2; \hat{Y}_1, \hat{Y}_2, \bar{Y}_1, \bar{Y}_2) \nonumber\\
    & = & h(\bar{Y}_1, \bar{Y}_2, \hat{Y}_1, \hat{Y}_2) -
    h(Z_1, Z_2) \nonumber\\
    & \leq & h(\bar{Y}_1, \bar{Y}_2) - h(Z_1, Z_2) + H(\hat{Y}_1, \hat{Y}_2) \nonumber\\
 & = & I(\bar{X}_1, \bar{X}_2; \bar{Y}_1, \bar{Y}_2) +
    H(\hat{Y}_1, \hat{Y}_2) \nonumber\\
   & \leq & I(\bar{X}_1, \bar{X}_2; \bar{Y}_1, \bar{Y}_2)
    + H(\hat{X}_1) + H(\hat{X}_2) , \label{eq:peak_power_mi}
\eea  where \eqref{eq:peak_power_mi} holds because $\hat{Y}_i$'s are
a function of $\hat{X}_i$'s from \eqref{eq:discard}.  Since $X_i$ is
$\mathcal{N}(0,1)$, we can show that \bean
    H(\hat{X}_i)
    \ = \ - \sum_{k \in \mathbb{Z}}
    p_{\hat{X}_i}(k) \log p_{\hat{X}_i}(k) \label{eq:quant_g}
    < \ 4. \label{eq:quant_g_end}
\eean  From \eqref{eq:peak_power_mi} and above, channel
\eqref{eq:quant} loses at most $8$ bits compared to channel
\eqref{eq:scale}.

Since $\hat{x}_i$ are not necessarily positive, we obtain positive
inputs by adding $(h_{i1}^{'} + h_{i2}^{'})$ to $\bar{y}_i$: \bea
    \nonumber \tilde{y}_i & := & \bar{y}_i + h_{i1}^{'} + h_{i2}^{'} \\
    & = & h_{i1} (\bar{x}_1 + 1)/2 + h_{i2} (\bar{x}_2 + 1)/2 + z_i
    \nonumber \\
    & =: & h_{i1} \tilde{x}_1 + h_{i2} \tilde{x}_2 + z_i, \ i = 1, 2,
    \label{eq:x_pos}
\eea where now $\tilde{x}_i$ lies in $[0,1)$.  $I(\tilde{X}_1,
\tilde{X}_2; \tilde{Y}_1, \tilde{Y}_2)$ remains equal to
$I(\bar{X}_1, \bar{X}_2; \bar{Y}_1, \bar{Y}_2)$.

The features of the model that we next address are
\begin{enumerate}
    \item channel gains are integers,
    \item inputs are restricted to \\
    $n \ := \ \max_{(i,j) \in \mathcal{E}} \lfloor \log h_{ij} \rfloor$ bits,
    \item there is no AWGN, and
    \item outputs involve truncation to integers.
\end{enumerate}  Let the binary expansion of $\tilde{x}_{i}$ be
$0.\tilde{x}_{i}(1) \tilde{x}_{i}(2) \ldots$.  We get the output of
the discrete superposition channel by retaining the relevant
portion of signal \eqref{eq:x_pos}: \bea
    y^{\text{det}}_i
    & =: & \left\lfloor \hat{h}_{i1} x^{\text{det}}_1 \right\rfloor
    + \left\lfloor \hat{h}_{i2} x^{\text{det}}_2 \right\rfloor ,\label{eq:final}
\eea where $x^{\text{det}}_i := \sum_{k = 1}^{n} \tilde{x}_i(k)
2^{-k}$, and we have truncated channel gains, i.e., $\hat{h}_{ij} :=
(\text{sign } h_{ij}) \lfloor |h_{ij}| \rfloor$.  To get
\eqref{eq:final} from \eqref{eq:x_pos}, we subtracted \bea
    \epsilon_i & := & \sum_{j = 1}^{2} \left(
    h_{ij} (\tilde{x}_j - x^{\text{det}}_j) +
    (h_{ij} - \hat{h}_{ij}) \, x^{\text{det}}_j \right. \nonumber \\
    & & \left. + \ \text{frac}\left(\hat{h}_{ij} \, x^{\text{det}}_j \right) \right) + z_i \label{eq:eps_dis} \\
    & =: & \sum_{j = 1}^{2} w_{ji} + z_i.
\eea  To bound the loss in mutual information, note \bean
    \lefteqn{I(\tilde{X}_1, \tilde{X}_2; \tilde{Y}_1, \tilde{Y}_2)} \nonumber\\
    & = & h(\tilde{Y}_1, \tilde{Y}_2) - h(\tilde{Y}_1, \tilde{Y}_2 | \tilde{X}_1, \tilde{X}_2) \nonumber \\
    & = & h(\tilde{Y}_1, \tilde{Y}_2) - h(Z_1) - h(Z_2) \nonumber \\
    & \leq & h(Y^{\text{det}}_1, Y^{\text{det}}_2, \epsilon_1, \epsilon_2) - h(Z_1) - h(Z_2) \nonumber   \\
    & \leq & H(Y^{\text{det}}_1, Y^{\text{det}}_2) + h(\epsilon_1) + h(\epsilon_2)  \\
    & & - \ h(Z_1) - h(Z_2). \nonumber
\eean  From the definition of $\epsilon_i$ in \eqref{eq:eps_dis}, and
since $y^{\text{det}}_i$ are completely determined by
$x^{\text{det}}_1, x^{\text{det}}_2$, we can rewrite \bean
    & & I(\tilde{X}_1, \tilde{X}_2; \tilde{Y}_1, \tilde{Y}_2)
    \  \leq \  I(X^{\text{det}}_1, X^{\text{det}}_2; Y^{\text{det}}_1,
    Y^{\text{det}}_2) \nonumber \\
    & & + \sum_{i = 1}^{2}(h(\epsilon_i) - h(\epsilon_i | W_{1i}, W_{2i})) \\
    & & \ = \ I(X^{\text{det}}_1, X^{\text{det}}_2; Y^{\text{det}}_1,
    Y^{\text{det}}_2) \\
    && + \ I(W_{11}, W_{21}; \epsilon_1)
    + I(W_{12}, W_{22}; \epsilon_2). \nonumber
\eean   By bounding the magnitudes of terms in \eqref{eq:eps_dis}, we
get $|w_{ji}| \ \leq \ 4$. So, $I(W_{11}, W_{2i};\epsilon_1)$ is the
mutual information of a MISO channel with input power constraint
$(16 + 16) = 32$ and $I(W_{11}, W_{21};\epsilon_1) \leq \frac{1}{2}
\log(1 + 32) < 3$ bits.  So we lose at most $6$ bits in the last
step.

We have proved that difference between the maximum mutual
information across a cut in a Gaussian network and an achievable
mutual information for the same cut in the discrete superposition
network is bounded.  Repeating this for every cut yields a bound
that depends solely on number of nodes.

Conversely, start with a joint distribution for inputs
$x^{\text{det}}_i$ in the discrete superposition network. Since
$x^{\text{det}}_i$'s satisfy average power constraint, we can apply
them directly to the Gaussian channel to get \beqn
    y_i \ = \ h_{i1} x^{\text{det}}_1 + h_{i2} x^{\text{det}}_2 + z_i,
    i = 1, 2.
\eeqn  We can rewrite $y_i$ as \bean
    {y_i} & = & y^{\text{det}}_i +  \sum_{j = 1}^{2} \left(
    (h_{ij} - \hat{h}_{ij}) \, x^{\text{det}}_j \right. \label{eq:eps_dis_2} \nonumber \\
    & & \left. + \ \text{frac}(\hat{h}_{ij} \ x^{\text{det}}_j
    ) \right) + z_i \nonumber \\
    & =: & y^{\text{det}}_i + \sum_{j = 1}^{2} v_{ji} + z_i , \ i = 1, 2.
    \label{eq:det_plus_noise}
\eean  By definition $y^{\text{det}}_i$ takes on only integer
values. Hence $y^{\text{det}}_i$ can be recovered from $y_i$, the
integer parts of $v_{ji}$'s and noise $z_i$, and the carry $c_i$
obtained from adding the fractional parts of $v_{ji}$'s and $z_i$.
So, \bean
    \lefteqn{I(X^{\text{det}}_1, X^{\text{det}}_2; Y^{\text{det}}_1,
    Y^{\text{det}}_2)} \nonumber\\
    & \leq & I(X^{\text{det}}_1, X^{\text{det}}_2; Y_1, Y_2,
    \{\hat{V}_{ji}\}, \{\hat{Z}_i\}, \{C_i\}) \\
    & \leq & I(X^{\text{det}}_1, X^{\text{det}}_2; Y_1, Y_2)
    + \sum_{i, j = 1}^{2} H(\hat{V}_{ji}) \nonumber\\
    & & + \sum_{i = 1}^{2} H(\hat{Z}_i) + \sum_{i = 1}^{2}
    H(C_i).
\eean Here $\hat{v}_{ji}$, $\hat{z}_i$ are integer parts of the
respective variables.  Since $|v_{ji}| \leq 2$, $\hat{v}_{ji} \in
\{-2, -1, \ldots, 2 \}$, and $H(\hat{V}_{ji}) \leq 3$. The carry
$c_i \in \{-2, -1, \ldots, 2\}$, hence $H(C_i) \leq 3$. As earlier,
$H(\hat{Z}_i) \leq 4$. Therefore mutual information of the
Gaussian channel is at most $28$ bits lesser than that of the
discrete superposition channel.

Above arguments can be extended to show that for every joint distribution for inputs in the discrete superposition
channel, there is a product distribution for inputs with a bounded
reduction in the mutual information. To complete the proof we note
that the cut-set bound in \eqref{eq:cs_bd} restricted to product
distributions is achievable in the discrete superposition
network \cite{ADT2}.

For the general case of complex Gaussian networks, we allow signals
in \eqref{eq:start} to be complex valued and rewrite \bean
    & & \left[ \begin{array}{c} y_{iR} \\ y_{iI}
    \end{array} \right] \ = \
    \left[ \begin{array}{cc} h_{i1R} & -h_{i1I} \\
    h_{i1I} & h_{i1R} \end{array} \right]
    \left[ \begin{array}{c} x_{1R} \\ x_{1I}
    \end{array} \right] \\
    & & + \left[ \begin{array}{cc} h_{i2R} & -h_{i2I} \\
    h_{i2I} & h_{i2R}  \end{array} \right]
    \left[ \begin{array}{c} x_{2R} \\ x_{2I}
    \end{array} \right]
    + \left[ \begin{array}{c} z_{R} \\ z_{I}
    \end{array} \right].
\eean  Now choose inputs to be i.i.d. $\mathcal{CN}(0,1)$.
Increasing variances of both inputs and Gaussian noise from $1$ to
$2$ does not change the mutual information, though
now $X_{iR}$, $X_{iI}$, $Z_{R}$, $Z_{I}$ are $\mathcal{N}(0,1)$. Rest of the analysis can be repeated.
\end{proof}

\section{MIMO channels and deterministic models} \label{sec:mimo}
Above we analyzed a network by comparing the MIMO channels
corresponding to the same cut in the Gaussian network and its
deterministic counterpart.  We can easily extend the negative result
in Sec.~\ref{sec:neg} to show that, in general, MIMO channels cannot
be approximated by the linear deterministic model.  In the same
vein, we can extend the positive result in Sec.~\ref{sec:dis_det} to
prove that the discrete superposition model remains a good
approximation for MIMO Gaussian channels.

\section{Concluding remarks}

Since the capacity of the linear deterministic model does not approximate that of the Gaussian relay network,
the challenge is to quantitatively show to what extent, how, and
why good {\em coding} strategies for the former yield good
strategies for the latter.

For discrete superposition networks, the challenge is to extend
the bounded error approximation result to networks with multiple
sources and destinations.

\section*{Acknowledgement}
The authors thank one of the reviewers for pointing out that, independently, results similar to Theorem~\ref{thm:det_gap} and the example in Section~\ref{sec:zero_phase} are obtained in \cite{AveThe}, though the approximating networks are different.  Specifically, in \cite{AveThe}, the inputs and channel gains for the approximating deterministic network are allowed to be complex valued while we restrict the inputs to a discrete set and the channel gains to points in $\mathbb{Z} + \imath \mathbb{Z}$ in order to simplify the model.


\begin{thebibliography}{1}

\bibitem{ADT1} A.~S.~Avestimehr, S.~N.~Diggavi, and D.~N.~C. Tse, ``A
Deterministic Approach to Wireless Relay Networks,'' {\em Proc.
45$^{th}$ Annu. Allerton Conf.}, Monticello, IL, Sept. 2007.

\bibitem{ADT2} -----------, ``Wireless
Network Information Flow,'' {\em Proc. 45$^{th}$ Annu. Allerton
Conf.}, Monticello, IL, Sept. 2007.

\bibitem{ADT3} -----------, ``Approximate
Capacity of Gaussian Relay Networks,'' {\em Proc. of
IEEE ISIT 2008}, Toronto, Canada, July 2008.

\bibitem{AST} A.~S.~Avestimehr, A.~Sezgin, and D.~N.~C. Tse,
``Approximate capacity of the two-way relay channel: A deterministic
approach,'' {\em Proc. 46$^{th}$ Annu. Allerton Conf.}, Monticello,
IL, Sep. 2008.

\bibitem{BT} G.~Bresler and D.~N.~C.~Tse, ``The two-user Gaussian interference
channel: a deterministic view,'' arXiv:0807.3222v1

\bibitem{CJS} V.~R.~Cadambe, S.~A.~Jafar, and S.~Shamai, ``Interference Alignment on
the Deterministic Channel and Application to Fully Connected AWGN
Interference Networks,'' accepted for publication in the {\em IEEE
Trans. on Info. Theory}. arXiv: 0711.2547.

\bibitem{Gam} A.~El~Gamal, ``On Information Flow in Relay Networks,'' {\em IEEE
National Telecomm. Conf.,} Vol.~2, pp.~D4.1.1~-~D4.1.4, Nov.~1981.

\bibitem{ETW} R.~Etkin, D.~N.~C. Tse, and H.~Wang,
``Gaussian Interference Channel Capacity to Within One Bit,'' arXiv:
0702045v2.

\bibitem{AveThe} A.~S.~Avestimehr, ``Wireless network information flow: a deterministic approach,'' Ph.D. dissertation, Univ of California, Berkeley, Oct. 2008.

\end{thebibliography}
\end{document}